\documentclass[twocolumn,amsmath,amssymb, preprintnumbers, showpacs, showkeys]{revtex4-1}

\usepackage{graphicx}
\usepackage{dcolumn}
\usepackage{bm}
\usepackage{amsmath, verbatim}

\begin{document}

\preprint{Version 5.0}

\title{Multi-particle Auger dissociation of excitons in shallow doped carbon nanotubes}
\author{Jay D. Sau${^1}$}%
\email{jaydsau@physics.harvard.edu}
\author{Jared J. Crochet${^{2}}$}%
\email{jcrochet@lanl.gov}
\author{Miguel Dimas${^2}$}%
\author{Juan G. Duque${^2}$}%
\author{Marvin L. Cohen${^3}$}%
\author{Stephen K. Doorn${^4}$}%

\affiliation{
${^1}$Department of Physics, Harvard University, Massachusetts\\
${^2}$Physical Chemistry and Applied Spectroscopy Group, Los Alamos National Laboratory, New Mexico\\
${^3}$Department of Physics, University of California at Berkeley and Materials Sciences Division, Lawrence Berkeley National Laboratory, Berkeley, California\\
${^4}$Center for Integrated Nanotechnologies, Los Alamos National Laboratory, New Mexico
}

\begin{abstract}
Shallow hole-doping in small diameter single-wall carbon nanotubes by H$_2$O$_2$ is shown to result in delocalized excited state quenching with no effects on the ground state absorption spectrum. To account for this process, the dissociation of excitons by shallow level electronic impurities is predicted to occur by multi-particle Auger decay. This mechanism, which relies on the chirality of the electronic states, causes the exciton to decay into electron-hole pairs with very high efficiencies.
  
\end{abstract}


\maketitle

For more than a decade, single-wall carbon nanotubes (SWNTs) have shown unique transport, optical, and mechanical properties making them one of the most promising systems for nanotechnology applications. The optical properties of single-walled carbon nanotubes, have proved to be particularly intriguing since they are dominated by strong resonances associated with bound excitons.  In particular the GW-Bethe Salpeter (GW-BSE) formalism \cite{hybertsen_louie, rohlfing_louie} for understanding the optical properties of semiconductors has been applied with great success to understanding the positions of the peaks in the absorption spectrum of SWNTs\cite{catalin_orig, georgy, ando}. Apart from the absorption frequency spectrum, the exciton lifetime also plays a critical role in determining the efficiency of  SWNTs as optical devices. As a result there have been several studies both theoretical\cite{catalin,paeie,mele} and experimental on the lifetimes of excitons\cite{Wang,Krauss}.

An examination of these studies reveals certain discrepancies between theory and experiment leading to the conclusion that the important problem of exciton decay is not understood. The most fundamental exciton lifetime mechanism for optically active excitons is radiative decay\cite{catalin}. However the calculated radiative decay  lifetime is found to be of the order of a few ns as opposed to experimentally measured lifetimes of a few tens of ps\cite{Berciaud2008}. Moreover, experimentally it is found that only about 1\% of the decay of excitons is radiative, ruling light emission out as a dominant mechanism\cite{Crochet2007}. In order to account for low emission efficiencies, SWNTs have been found to be hole doped spontaneously by the presence of the aqueous oxygen redox couple\cite{Aguirre2009}, molecular oxygen\cite{Yoshikawa2010}, and other redox species such as H$_2$O$_2$\cite{Crochet2011}. Therefore alternative decay mechanisms involving an unintentional hole doping such as Auger decay \cite{mele} and phonon based Auger decay\cite{paeie} have been proposed.

\begin{figure}
\includegraphics{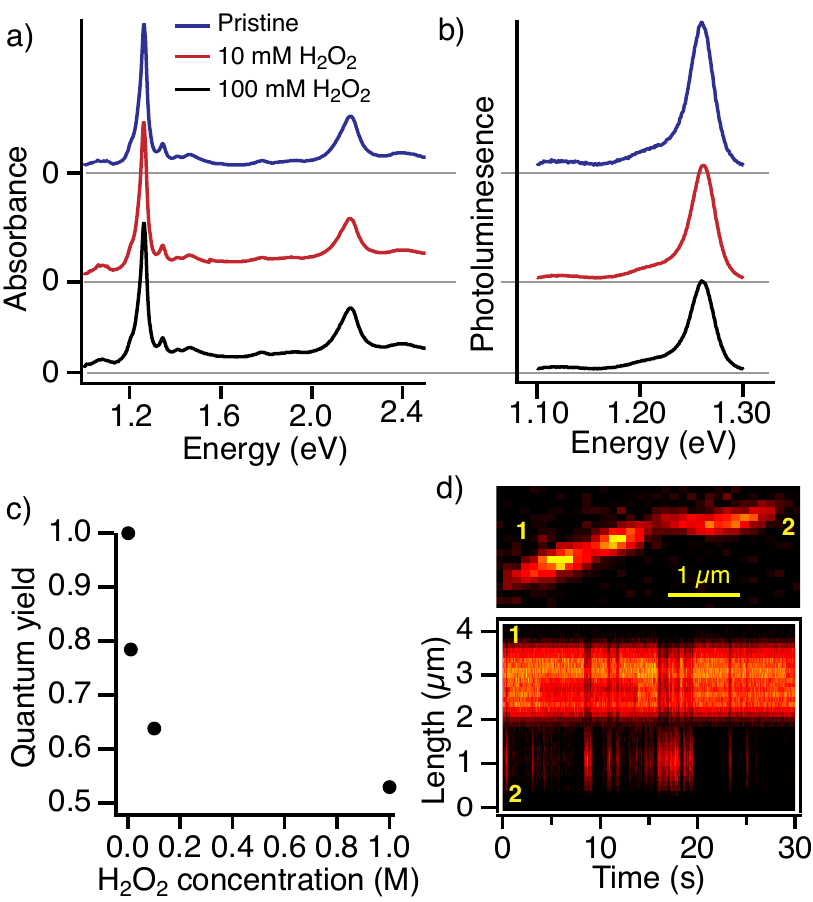}
\caption{a) and b) Absorption and photoluminescence spectra of an ensemble of (6,5) enriched SWNTs before and after the addition of H$_2$O$_2$. The spectra are unscaled raw data. c) Relative ensemble quantum yield versus concentration of H$_2$O$_2$. d) Photoluminescence image of an $\approx$ 4 $\mu$m (6,5) nanotube under 561 nm continuous wave excitation. Spatio-temporal photoluminescence intensity after the addition of 2 mM of H$_2$O$_2$ is shown in the bottom panel. Long range quenching of excitons is observed on separate regions of the tube. For more examples and details see Ref.\cite{Crochet2011}.}
\end{figure}
 
In this letter we first present experimental results of delocalized exciton quenching in colloidal (6,5) SWNTs by the common oxidant H$_2$O$_2$. This doping process results in intermittent photoluminescence (PL), while the absorption spectrum is unchanged. The strong quenching of emission leads to the conclusion that the dopant induced decay rate of excitons is much faster than that predicted by all current theories of exciton decay such as Auger decay and phonon assisted Auger decay. Therefore, we propose a new four particle Auger mechanism for the decay of excitons, which is purely a result of the Coulomb interaction. The proposed mechanism is unlike the conventional Auger mechanism, and leads to the generation of four particles (\textit{i.e.} two electron hole pairs) instead of a single high-energy hole. This eliminates the kinetic energy barrier involved in the Auger decay and allows the process to occur for zero momentum bright excitons. The rate of decay from this mechanism is calculated using a two band model with parameters fit to match first-principles calculations of excitons. The results are found to be in good agreement with delocalized emission quenching in the absence of bleaching of the absorbance. 

SWNTs synthesized by the HiPco method (batch $\#$ 187.4) were suspended in 1$\%$ aqueous deoxycholate by shear mixing for one hour.  Un-solubilized material was removed from the suspension by bench-top centrifugation and the resulting supernatant was structurally sorted by density gradient ultracentrifugation\cite{Ghosh2010}. Fractions enriched in the (6,5) tube were collected for spectroscopy and microscopy. For microscopy, the natural Brownian motion associated with SWNTs in solution was reduced for quenching experiments by adding a 10 $\mu$L drop of suspension on a plasma cleaned microscope cover glass and covering it with a second smaller cover glass that was sealed on three sides with vacuum grease. The spreading of the drop aided in the weak adhesion of tubes to the glass surface with a thin solution layer present and the tubes showed bright and stable luminescence\cite{Crochet2011}. Quenching experiments were carried out by adding a drop of reagent to the unsealed side of the cover glass where capillary forces facilitated rapid mixing or to bulk solutions. PL imaging was performed with an inverted microscope equipped with an electron-multiplying CCD camera (Princeton Instruments ProEm) and a 1.49 NA 60x objective where the total photon collection efficiency at 980 nm was estimated to be approximately 1.3$\%$. The excitation source consisted of a continuous-wave solid-state 561 nm laser diode where the photon flux density was kept at 7$\times$10$^{20}$ cm$^{-2}$s$^{-1}$. 

In Figs. 1a and b we show representative absorption and PL spectra of (6,5) enriched tubes before and after the addition of H$_2$O$_2$. No sign of bleaching of the ground state $S_1$ exciton (change in absorption strength) was visible. However, as shown in Fig. 1c, the quantum yield was reduced by varying the H$_2$O$_2$ concentration. The negligible effect on the ground state absorption is understood from a strong diameter dependence of the bleaching of $S_1$ by H$_2$O$_2$\cite{Song2005}. Such effects may arise from a diameter dependent valence band maximum, which underlies the absolute potential of the SWNT/surfactant system\cite{O'Connell2005}. Nonetheless, the emission at the ensemble and the single tube levels is affected and we have found that long range intermittent exciton quenching occurs at low reagent concentration regime, Figs. 1b and 1c. We have previously assigned this effect to shallow hole doping at the low dopant concentration level which can lead to exciton dissociation\cite{Crochet2011}. 

In order to determine the effects of shallow hole doping on the optical properties of SWNTs we compute the hole density in the valence band as a function of the Fermi energy from $n_d(E_F)=\int_{E_g/2}^{\infty} dE g(E) f(E, E_F)$, where the density of states of the valence band is approximated as,

\begin{equation}
g(E)=\frac{4}{\pi \hbar v_F}\frac{E}{\sqrt{E^2-(E_g/2)^2}},
\end{equation}

\noindent and $f(E, E_F)$ is the Fermi-Dirac distribution. Assuming a one-to-one reaction, $n_d$ would reflect the dopant density. Here the Fermi velocity is taken to be $v_F=1$ nm/fs. We estimated the normalized absorption strength as a function of $n_d$ using $A=1-n_d/n_0$ where $n_0$ is the number of holes required to perfectly bleach $S_1$\cite{Matsuda2011}. Using $n_0=4 (2 m_h E_b)^{1/2}/(\pi \hbar )$, where the hole mass is $m_h=0.15 m_e$\cite{Jorio2005}, the exciton binding energy is  $E_b=$350 meV, and a gap energy of $E_g=1.6$ eV, we find a weak dependence of the absorption strength on $E_F$ up until $E_F\ge E_g/2$. Therefore, we consider an excited state quenching mechanism in a partially un-occupied valence band.

\begin{figure}
\includegraphics{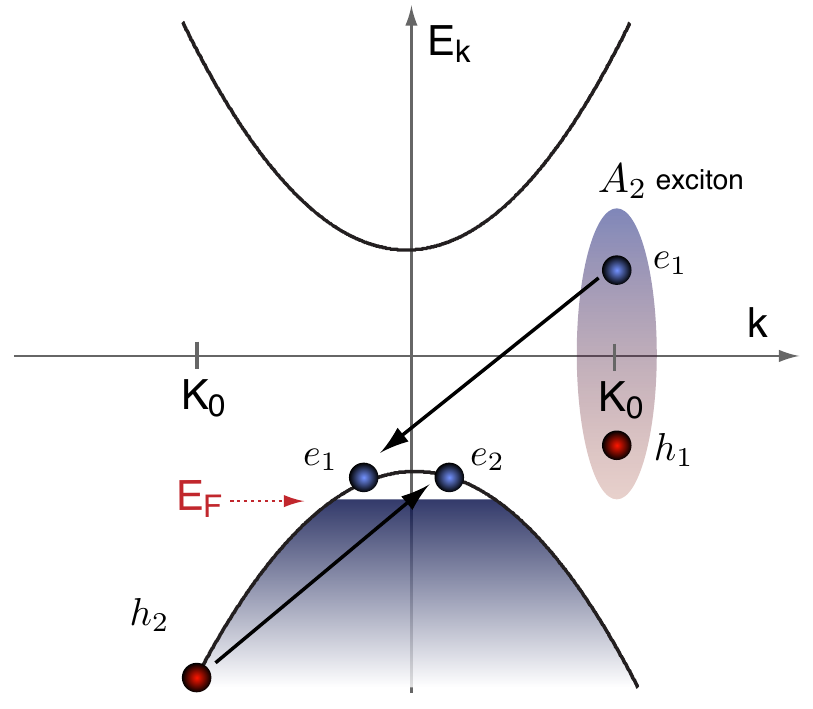}
\caption{Schematic of the bright $A_2$ symmetry exciton in a slightly doped single-walled carbon nanotube with a hole pocket in the otherwise filled valence band. The filled part of the valence band is shaded in blue. The exciton is composed of the vertically separated electron-hole pair on the extreme right. The black arrows show the proposed decay mechanism for the exciton. In this process the electron from the exciton $e_1$ make a transition to the hole pocket in the valence band. At the same time another electron $e_2$ makes a transition from the filled valence state to the hole pocket to conserve energy and momentum.}
\label{Fig2}
\end{figure}

In hole doped SWNTs photoexcitations can decay by the Auger process where the exciton decays by transferring its energy and momentum to a hole\cite{mele}. However this mechanism requires the exciton dispersion to intersect with the hole dispersion at an energy comparable to the thermal energy so that both energy and momentum can be conserved. This only leads to an efficient decay mechanism of dark excitons\cite{mele} which have a larger effective mass than the holes. Therefore the conventional Auger mechanism does not lead to decay of the zero momentum  bright excitons that are created by the incident radiation. The other mechanism that has been suggested for the decay of excitons in doped carbon nanotubes is the phonon assisted Auger (PAIEI) mechanism\cite{paeie}.  Here the phonon provides the excess momentum needed for the Auger decay. However, for low dopant dopant densities the PAEIE mechanism leads to a lifetime which is too long to be consistent with experiment\cite{Crochet2011}.

Given that the currently studied mechanisms are unable to account satisfactorily for the exciton lifetime, we consider a four particle Auger mechanism (shown in Fig. 2), where the electron from a $k=K_0$ exciton makes a transition to the hole pocket (which is allowed because of the chirality of the nanotube states), while to conserve energy and momentum another electron makes a transition from the occupied valence band into the hole pocket. Th chirality of the electronic states arises from the mirror symmetry of the $K/K'$ valleys associated with the underlying graphene sub-lattices which results in energetically degenerate, but opposite angular momentum states on the quantized tube. The resulting process is a decay of the exciton into a pair of electrons in the hole pocket together with a pair of holes in the occupied valence band. The interaction driving this scattering process is the screened Coulomb interaction of the semiconducting tube\cite{takeshima}. We are assuming that we are at sufficiently low doping so that the metallic screening can be ignored\cite{newcatalin}. Furthermore, we  take the Coulomb interaction to be in the static limit, and the calculated decay rate of the exciton is related to the strength of the screened Coulomb interaction which in turn will be estimated from the measured exciton energy.

The initial exciton state with center of mass momentum $K_0$ in the single band approximation can be written as  

\begin{align}
&|\Psi_{exc,K_0}\rangle=\int dk d\bm r_e d\bm r_hA_{cvk} \phi_{ck+\frac{K_0}{2}}(\bm r_e)\phi_{vk-\frac{K_0}{2}}^*(\bm r_h)
\nonumber\\
&c^\dagger (\bm r_e)c(\bm r_h)|0\rangle
\end{align}

\noindent where $\phi_{(c,v)k}$ are the conduction and valence band wave-functions for the carbon nanotube and $c^\dagger$ is the electron creation operator on the lattice of carbon atoms on the nanotube acting on the vacuum state $|0\rangle$\cite{rodrigo}. 
The band wave-functions $\phi$, the exciton wave-function $A_{cvk}$, and the exciton energy $\Omega_S$ have been obtained previously from first principles calculations \cite{hybertsen_louie, rohlfing_louie, catalin_orig}. The exciton decay  into  multi-quasiparticle states is obtained by considering the Coulomb interaction perturbatively on the exciton. The lowest order Coulomb interaction processes where an exciton decays into hole quasiparticles are shown in Fig. 2. To define the final decay state, we define the vacuum state $|0\rangle$ to be the state of the electrons with the hole pocket. In terms of the state $|0\rangle$, the final state with a pair of electrons at $k_{e_1},k_{e_2}$, both in the hole pocket in the valence band and a pair of holes at momentum $k_{h_1}$,$k_{h_2}$ in the occupied part of the valence band can be written as 

\begin{align}
&|k_{e_1},k_{e_2},k_{h_1},k_{h_2}\rangle=\int d\bm r_{e_1}d\bm r_{e_2}d\bm r_{h_1}d\bm r_{h_2}  \phi_{v k_{e_1}}(\bm r_{e_1})\nonumber\\
&\phi_{v k_{e_2}}(\bm r_{e_2})\phi_{vk_{h_1}}^*(\bm r_{h_1})\phi_{vk_{h_2}}^*(\bm r_{h_2})c^\dagger(\bm r_{e_1})c^\dagger(\bm r_{e_2})c(\bm r_{h_1})
\nonumber\\
&c(\bm r_{h_2})|0\rangle.
\end{align}

\noindent The final multiple quasiparticle states must continue to have a center of mass momentum
 $k_{e_1}+k_{e_2}-k_{h_1}-k_{h_2}=K_0$ and also total energy equal to the excitation energy
 $\epsilon_{v}(k_{e_1})+\epsilon_v(k_{e_2})-\epsilon_v(k_{h_1})-\epsilon_v(k_{h_2})=\Omega_S$ of the exciton.
 Here $\epsilon_v(k)$ is the valence band dispersion of the highest valence band of the nanotube. Neglecting the
 Fermi-energies of the hole-pocket (i.e. $|\epsilon_v(k_{e_1})|,|\epsilon_v(k_{e_2})|\ll\Omega_S$), the final hole
 energies can be approximated by $\Omega_S\approx -(\epsilon_v(k_{h_1})+\epsilon_v(k_{h_2}))$.  Since the thermal energy
 of the exciton is much lower than $\Omega_S$ (so that $K_0\ll |k_{h_1}|,|k_{h_2}|$), we can further approximate
 $\epsilon_v(k_{h_1})\approx \epsilon_v(k_{h_2})\approx -\Omega_S/2$ and $k_{h_1}\approx -k_{h_2}$.
 Furthermore from Fig. 2 it is clear that one of the holes remains at exactly the same momentum as it was in the exciton.
 Therefore there are 4 processes that are related by particle exchange to obtain the same final state starting from the exciton.
 All these processes are mediated by the screened Coulomb interaction\cite{takeshima}. However, since $k_{e_1}\approx k_{e_2}$
 are in the hole pocket and they are small, the direct and exchange contributions from these processes cancel to suppress
 the process where all e-h pairs are in the same band. Therefore we consider the case where the excitons scatters into
 an e-h pair in the same pair of bands, while the other e-h pair is created in a different pair of bands.
 In that case, we can assume that $k_{e_1}$ and $k_{h_1}$ arise from the exciton and the Auger process creates
 the pair $k_{e_2}$ and $k_{h_2}$. The result is a single process that contributes to the amplitude and is given by 

\begin{equation}
Q(K_0;k_{e_2},k_{h_2})\approx A_{cv k_{h_2}}W(k_{h_2}+K_0,k_{h_2};k_{h_2})
\end{equation} 

\noindent where we have approximated  $k_{e_2}\approx 0$. In the above 

\begin{align}
&W(k_1,k_2;q)=\int d\bm r_1 \int d\bm r_2  \phi_{c,k_1}(\bm r_1)\phi_{v,k_2}(\bm r_2)\nonumber\\
&\phi_{v,k_2+q}^*(\bm r_2)\phi^*_{v,k_1-q}(\bm r_1) W(\bm r_1;\bm r_2)
\end{align}

\noindent and $W$ is the static screened Coulomb interaction. Assuming the hole pocket to be at temperature
 $T\ll\Omega_S/k_B$, the exciton decay rate will be given by 

\begin{equation}
k_{Aug.}=\frac{2\pi}{\hbar} Q^2 n_d^2 \label{eq:rate1}.
\end{equation}

Given the small doping levels of the nanotube, we will restrict ourselves to a simplistic two-band continuum model
\cite{ando} to estimate the decay rate. Within this approximation, the screened electron-hole interaction is taken
 to be of the form $U(\frac{q}{k_x})=g I_0(|\frac{q}{k_x}|)K_0(|\frac{q}{k_x}|)$ where $I_0$ and $K_0$ are Bessel functions.
 Here $k_x=\frac{2}{3 d}$ is the transverse momentum quantum for the electrons on the nanotube and $d$ is the diameter of
 the nanotube. The two band continuum BSE that determines the exciton wave-function can be written in dimensionless form as

\begin{align}
&\Omega(k)=(\sqrt{1+k^2}-\varepsilon)B(k)-\lambda\int dq U(q)F(k;k+q)
\nonumber\\
&B(k+q)=0   
\end{align}

\noindent where $\lambda=g k_x/{E_g}$, $\varepsilon=\Omega_S/{E_g}$, and $E_g=2 \hbar v_F  k_x$ is the gap of
 the nanotube within the $k\cdot p$ approximation. Here, $v_F$ is the Fermi velocity and the chirality factor
 $F(k_1;k_2)=|\langle \phi_{c,k_1}|\phi_{c,k_2}\rangle|=|\langle \phi_{v,k_1}|\phi_{v,k_2}\rangle|$.
  Here the dimensionless solution $B(k)$ to the exciton wave-function is $A(q)=B(\frac{q}{k_x})/\sqrt{k_x}$.
 Previous calculations for the screened Coulomb interactions in nanotubes\cite{paeie} have shown that
 $B$ scales as $B(k)\approx\sqrt{0.77 r_0 k_x}/(1+k^2 r_0^2 k_x^2)^{1.3}$ where $r_0\approx 1.95 d$.
 Using this form for the wave-function $B(k)$, we can optimize the dimensionless parameters $\varepsilon$ and
 $\lambda$ to make $B(k)$ fit the BSE best by minimizing $\int dk \Omega(k)^2$. The result is that we find a gap
 scaling as  $\Omega_S= 0.506 E_g$ and $g=0.233 E_g/k_x$.

\begin{figure}
\includegraphics{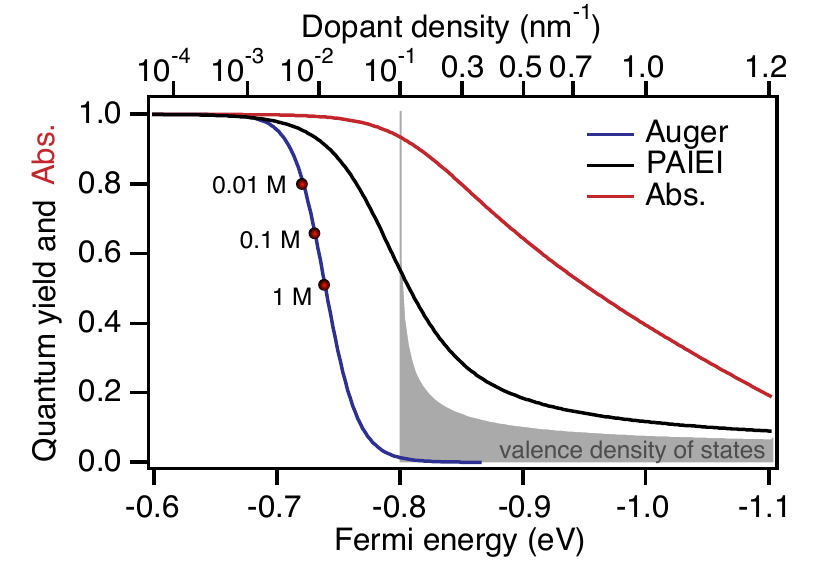}
\caption{Calculated normalized absorption strength (red) and relative fluorescence quantum yields $\eta$ for the PAIEI (black)\cite{paeie} and multi-particle Auger mechanisms (blue), Eq. \ref{eq:central}, as a function of Fermi energy. The density of states of the valence band is shown in grey and experimental quantum yields from Fig. 1 are given by the red circles for comparison.}
\label{Fig2}
\end{figure}

Given the coupling constant $g$ the screened Coulomb interaction kernel $W$ can be determined from
 $U$ using the Bloch functions $\phi_{(c,v),k}$ within the $k\cdot p$ approximation\cite{ando}. One can then 
simplify Eq.~\ref{eq:rate1} to the relation

\begin{equation}
k_{Aug.}=3.5\times 10^{-2}\frac{k_F^2}{k_x^2}\frac{E_g}{\hbar}.
\end{equation}

\noindent The Fermi momentum $k_F$ is given by the number of hole dopants per unit tube length or in terms of the
 dopant density $n_d$ such that $k_F=\pi n_d/4$ where the factor of four is from the spin and $K/K'$ valley degeneracies.
 This leads to the final equation for the non-radiative decay in ps$^{-1}$, 

\begin{equation}
k_{Aug.}\approx 100 n_d^2 d.\label{eq:central} 
\end{equation}
This equation is the central result of our analysis for the 4-particle Auger decay in carbon nanotubes.

Using Eq. 9 with the standard definition of the fluorescence quantum yield, 
\begin{equation}
\eta=k_r/(k_r+k_{nr}+k_{Aug.})\label{eq:eta}
\end{equation}
 where $k_r$ is
 the radiative decay rate and $k_{nr}$ is the non-radiative rate before hole doping, we show a strong change in $\eta$ as a function of the Fermi level in Fig. 3 at room temperature for multi-particle Auger decay. Here $k_r=7\times 10^{-4}$ ps$^{-1}$ and $k_{nr}=1\times10^{-2}$ ps$^{-1}$\cite{Hertel2010, Gokus2010}.  We also compared the PAIEI mechanism\cite{paeie} for the same parameters used above and find our mechanism proposed here is much more efficient, Fig. 3. Therefore, this model suggests that the multi-particle Auger decay can be the dominant mechanism of excitonic decay in shallow doped small diameter carbon nanotubes even when bleaching of the absorption strength is negligible.

In summary, we have demonstrated that excitons can be effectively dissociated in small diameter single-wall carbon nanotubes even at relatively low dopant concentrations. This extremely efficient mechanism is driven by both the particularly strong Coulomb interaction experienced by charge carriers in quasi one-dimensional systems and the intrinsic chirality of the nanotube lattice. Also, these results can account for the exponential quenching of SWNT emission in field effect transistor devices as a function of gate voltage\cite{Yasukochi2011}. Finally, we emphasize the potential of carbon nanotubes as active elements in photovoltaic devices where simultaneous strong light absorption and exciton dissociation is desired.

J.S. thanks the Harvard Quantum Optics Center, JQI-NSF-PFC, DARPA-QUEST and LPS-NSA and M.L.C. acknowledges the
 NSF grant DMR07-05941 and Director, Office of Science, Basic Energy Sciences, Materials Sciences and Engineering Division 
of the U.S. Department of Energy under contract No. DE-AC02-05CH11231. This work was performed, in part, at the Center for
 Integrated Nanotechnologies, a U.S. Department of Energy, Office of Basic Energy Sciences user facility.  Los Alamos
 National Laboratory is operated by Los Alamos National Security, LLC, for the National Nuclear Security Administration
 of the U.S. Department of Energy under contract DE-AC52-06NA25396.

\end{document}